\documentclass[letterpaper]{article} 
\usepackage{aaai25}  
\usepackage{times}  
\usepackage{helvet}  
\usepackage{courier}  
\usepackage[hyphens]{url}  
\usepackage{graphicx} 
\usepackage{natbib}  
\usepackage{caption} 
\usepackage{algorithm}
\usepackage{algorithmic}

\usepackage{newfloat}
\usepackage{listings}
\DeclareCaptionStyle{ruled}{labelfont=normalfont,labelsep=colon,strut=off} 
\floatstyle{ruled}
\newfloat{listing}{tb}{lst}{}
\floatname{listing}{Listing}
\nocopyright

\usepackage{multirow}
\usepackage{multicol}
\usepackage{amsfonts}
\usepackage{subfigure}
\usepackage{diagbox}
\usepackage{booktabs}
\usepackage{arydshln}
\usepackage{bbding}
\usepackage[most]{tcolorbox}
\usepackage{pifont}

\graphicspath{{./}{Figures/}}
\newcommand{\tabincell}[2]{\begin{tabular}{@{}#1@{}}#2\end{tabular}}

\newcommand{\colorred}[1]{#1}
\newcommand{\colorblue}[1]{#1}

\usepackage{booktabs}
\usepackage{multirow}
\usepackage{amsmath}
\usepackage{amsfonts}
\usepackage{pifont}

\title{Lightweight Joint Audio-Visual Deepfake Detection  via \\ Single-Stream Multi-Modal Learning Framework}

\author{
    Kuiyuan Zhang\textsuperscript{\rm 1},
    Wenjie Pei\textsuperscript{\rm 1},
    Rushi Lan\textsuperscript{\rm 2},
    Yifang Guo\textsuperscript{\rm 3},
    Zhongyun Hua\textsuperscript{\rm 1}\thanks{Corresponding author},
}
\affiliations{
    \textsuperscript{\rm 1} Computer Science and Technology, Harbin Institute of Technology Shenzhen, Shenzhen, China\\
    \textsuperscript{\rm 2} Computer Science and Information Security, Guilin University of Electronic Technology, Guilin, China\\
    \textsuperscript{\rm 3} Alibaba Group, Hangzhou, China\\

    zkyhitsz@gmail.com, wenjiecoder@outlook.com, rslan2016@163.com, guoyifang@gmail.com, huazyum@gmail.com
}

\usepackage{bibentry}

\begin{document}

\maketitle

\begin{abstract}
Deepfakes are AI-synthesized multimedia data that may be abused for spreading misinformation. 
Deepfake generation involves both visual and audio manipulation. 
To detect audio-visual deepfakes, previous studies commonly employ two relatively independent sub-models to learn audio and visual features, respectively, and fuse them subsequently for deepfake detection. However, this may underutilize the inherent correlations between audio and visual features. Moreover, utilizing two isolated feature learning sub-models can result in redundant neural layers, 
making the overall model inefficient and impractical for resource-constrained environments.
 In this work, we design a lightweight network for audio-visual deepfake detection via a single-stream multi-modal learning framework. 
Specifically, we introduce a collaborative audio-visual learning block to efficiently integrate multi-modal information while learning the visual and audio features. By iteratively employing this block, our single-stream network achieves a continuous fusion of multi-modal features across its layers. Thus, our network efficiently captures visual and audio features without the need for excessive block stacking, resulting in a lightweight network design. 
Furthermore, we propose a multi-modal classification module that can boost the dependence of the visual and audio classifiers on modality content. It also enhances the whole resistance of the video classifier against the mismatches between audio and visual modalities. We conduct experiments on the DF-TIMIT, FakeAVCeleb, and DFDC benchmark datasets. Compared to state-of-the-art audio-visual joint detection methods, our method is significantly lightweight with only 0.48M parameters,
yet it achieves superiority in both uni-modal and multi-modal deepfakes, as well as in unseen types of deepfakes. 
\end{abstract}

\begin{figure}[!htbp]
    \centering
    \includegraphics[width=0.9\linewidth]{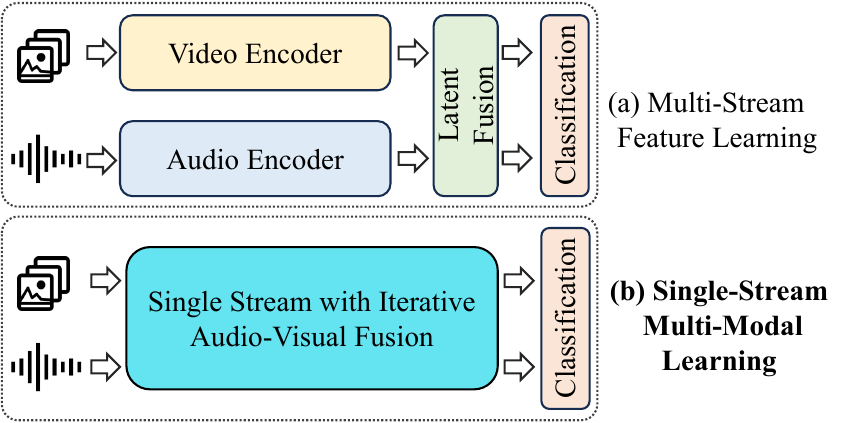}
    \caption{Illustration of two feature learning frameworks in audio-visual deepfake detection. The multi-stream feature learning framework only fuses visual and audio features before classification, leading to the under-utilization of multi-modal features. In contrast, our single-stream multi-modal learning framework can iteratively fuse multi-modal information throughout the feature learning process.}
    \label{fig:acc-speed}
\end{figure}

\section{Introduction}
The development of computer vision and deep learning has driven the creation of increasingly realistic visual content through visual generation methods~\cite{Aynur2022-deepfake_review,Siwei2020-deepfake_review}, which are commonly referred to as ``deepfakes''. Initially, deepfakes only involved the visual modality due to the limited capabilities of audio generation methods. However, recent advances in text-to-speech (TTS) and voice conversion (VC) methods have shown significant improvement in audio generation quality~\cite{kim2021conditional-VITS,casanova2022yourtts}. The improved quality of visual and audio deepfakes has made it more flexible to combine or generate desired videos and audios. In real scenarios, video deepfakes can involve the forgery of the visual, audio or both modalities. As many open-source deepfake generation methods become increasingly accessible and easy to use, it should pay more attention to the potential security threats of multi-modal deepfakes.

Many researchers are devoted to the research of deepfake detection by either constructing benchmark datasets~\cite{kwon2021kodf,li2020celebdf,huang2021deepfake-deepfakeMNIST} or proposing deepfake detection methods~\cite{li2022waveletenhanced,he2022defeatingdeepfake}. However, previous works mainly concern uni-modal detection that verifies whether a visual frame, a visual clip or an audio clip is fake or real. These methods rely on specific cues of a single modality, which can be repaired using adversarial training or more advanced generation methods~\cite{yang2022deepfake-architecture-attribution}. Besides, a video in a real scenario may be forged on both visual and audio modalities~\cite{cai2022you-LAVDF}. To detect whether a single modality or the whole video is forged, one must combine multiple uni-modal detection methods to classify the video, e.g., ensemble learning. This strategy may cause extra computational resources. Besides, it ignores the correlations between visual and audio modalities. The video may be misclassified when its visual and audio modalities are carefully synthesized.

Recently, researchers have developed several audio-visual joint detection networks that can utilize both visual and audio modalities for deepfake detection. They combine the multi-modal information to detect whether the whole video has been manipulated or not~\cite{mittal2020emotions-EmotionsDontLie,cheng2022voice_VFD}. Some methods can also detect that either the visual or audio modality has been manipulated~\cite{zhou2021joint-2+1,cai2022you-LAVDF,razaMultimodaltraceDeepfakeDetection2023}. These multi-modal detection methods have the ability to utilize the potential correlations and consistencies inherent in visual and audio content, leading to high detection accuracy. However, these methods tend to learn visual and audio features independently, only fusing them during the final classification stage. Unfortunately, this strategy overlooks the potential correlations between these modalities during the feature learning process. Moreover, utilizing two isolated feature learning sub-models can result in redundant neural layers. As a result, they commonly adopt complex network architectures containing numerous parameters, rendering them unsuitable for deployment on some resource-limited devices, such as mobile devices.



In this paper, we propose a lightweight \textbf{S}ingle-\textbf{S}tream network for joint \textbf{A}udio-\textbf{V}isual deepfake \textbf{D}etection called \textbf{SS-AVD}.
Specifically, we first design a \textbf{c}ollaborative \textbf{a}udio-\textbf{v}isual \textbf{l}earning (CAVL) block to learn the visual and audio features interactively. 
The CAVL block consists of a visual preprocessing module (VPM) and a self-attention-based audio-visual module (SAAVM). VPM utilizes spatial attention to capture the spatial dependencies of each visual frame, while SAAVM learns the temporal-spatial correlations between the visual and audio modalities using the attention mechanism. 
By stacking CAVL blocks, we build a lightweight single-stream network with iterative feature fusion.
Furthermore, we propose a multi-modal classification module to detect the visual, audio, and the whole video. The module comprises two main strategies: multi-modal style-shuffle augmentation (MMSSA) strategy and latent-shuffle augmentation (LSA) strategy. The MMSSA strategy randomly shuffles the styles of the latent features for each modality, enabling the visual and audio classifiers to rely more on the feature content. The LSA strategy randomly combines the visual and audio features from different samples to enhance the classifier's robustness against potential mismatches between the visual and audio modalities. Finally, we conduct extensive experiments on three audio-visual benchmark datasets
to evaluate our SS-AVD.

The main contributions of our work are summarized as follows:
\begin{itemize}
    \item We develop a lightweight audio-visual joint detection model via a single-stream multi-modal learning framework. Differing from existing methods that utilize two independent sub-models to individually learn audio and visual features, our approach fuses multi-modal information while learning the visual and audio features. As a result, our network can efficiently capture visual and audio features without the need for excessive block stacking, resulting in a lightweight network design. 
    


    \item  We propose a multi-modal classification module that can boost the dependence of the visual and audio classifiers on modality content. It also enhances the whole resistance of the video classifier against the mismatches between audio and visual modalities.

    \item  Extensive experiments show that our method is significantly lightweight with only 0.48M parameters (usually greater than 5M in previous methods),  
    yet it achieves superiority in both uni-modal and multi-modal deepfakes, as well as in unseen types of deepfakes, compared to state-of-the-art audio-visual joint detection methods.

\end{itemize}

\section{Related work}
This section first introduces the deepfake generation and further presents the uni-modal and multi-modal deepfake detection.

\subsection{Deepfake Generation}
Deepfake in the early stage mainly focuses on the forgery of visual content. The term ``deepfake'' was first proposed by a user of Reddit who used a face-swap method to replace faces in videos~\cite{David-deepfake_redit}. Currently, the development of visual deepfake technology has made the generated fake images and videos increasingly close to the real data~\cite{groshev2022ghost}. 

The audio deepfakes are usually the cloned voice of a person in a video or the generated waveform according to a text. TTS~\cite{kim2021conditional-VITS} and VC~\cite{wang2022drvc} are two commonly used techniques for creating audio deepfakes~\cite{frank2021wavefake}. TTS synthesizes audio from the input text, while VC converts the rhythm, timbre, or pitch of a voice from a person to make the voice sound like another person. With the assistance of deep learning, TTS can now synthesize very clearly but mechanical audio, while VC can generate natural audio that is hard to recognize by human beings.

A video in real scenarios may be forged on the visual, audio, or both modalities. To synthesize a high-quality video, some lip-syncing works, such as AttnWav2Lip~\cite{wang2022attentionbased-Lipsync},  have been proposed to synchronize the visual modality with the audio modality such that the video is more natural for human beings.


\begin{figure*}[!htbp]
    \centering
    \includegraphics[width=0.95\linewidth]{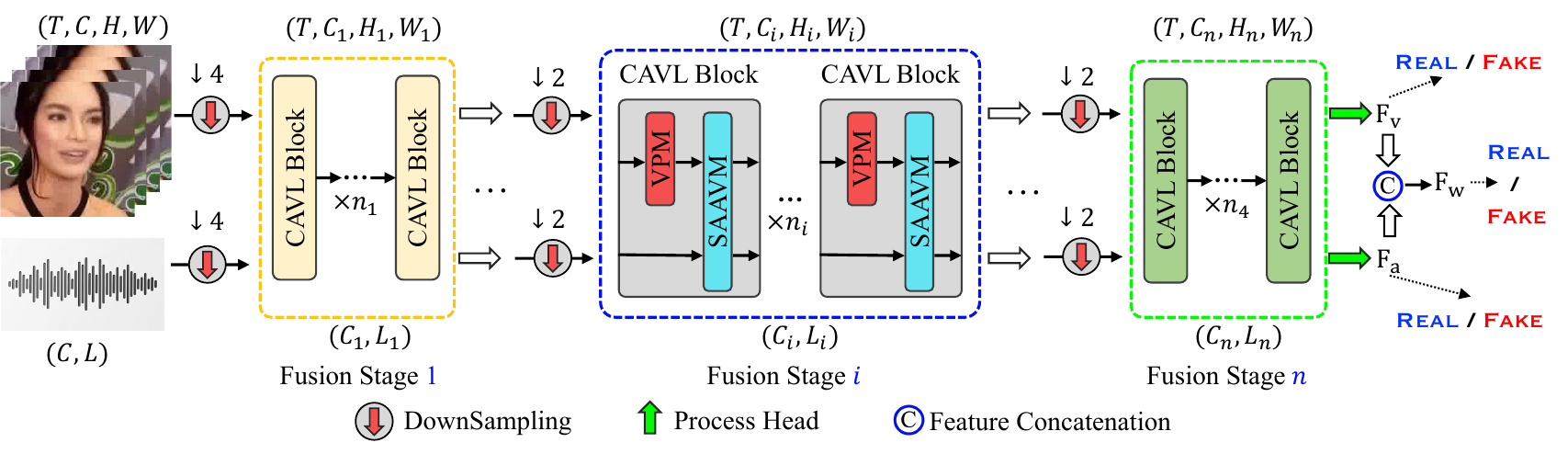}
    \caption{Overview of our SS-AVD. Given the input visual and audio clips, our SS-AVD sequentially fuses the visual and audio features at multi-scale using $n$ fusion stages, which are constructed by employing stacked CAVL blocks. Finally, our SS-AVD simultaneously predicts the labels for the visual, audio, and video.}
    \label{fig:architecture}
\end{figure*}

\subsection{Uni-modal DeepFake Detection}
Previous deepfake detection methods can detect only video frames or video clips~\cite{wang2020CNNAug}. 
Some classical image classification models, such as Xception~\cite{chollet2017xception} and EfficientNet~\cite{tan2019efficientnet}, have been applied to frame-based deepfake detection. Since these models can learn high-level semantic features, they have shown significant performance in deepfake detection tasks. Besides, deepfakes usually contain synthetic features that are different from natural images. Some works improve the detection accuracy utilizing these unique characteristics, such as the abnormal frequency features~\cite{jeong2022bihpf}, identity inconsistency~\cite{dong2022protecting}, image consistency~\cite{zhao2021learning_self-consistency}, and the trace of generation models~\cite{yang2022deepfake-architecture-attribution}. Though frame-based detection methods can average the classification scores of each video frame to detect video clips, they cannot utilize the inter-frame correlation.

The video deepfake detection methods can exploit the inter-frame and intra-frame features of a video clip simultaneously~\cite{haliassos2021-LipForensics,zhao2023istvt}. For example, Haliassos $et~al.$~\cite{haliassos2021-LipForensics} proposed LipForensics for deepfake video detection, which is pre-trained on lipreading in the wild dataset to learn the latent features of lip movement and utilizes the disharmonious lip movements to detect video deepfake. 

These uni-modal deepfake detection methods can only detect visual deepfakes and cannot detect videos with possible forged audio. This limits their applicability to real scenarios.

\begin{figure}[!tbp]
    \centering
    \includegraphics[width=1.0\linewidth]{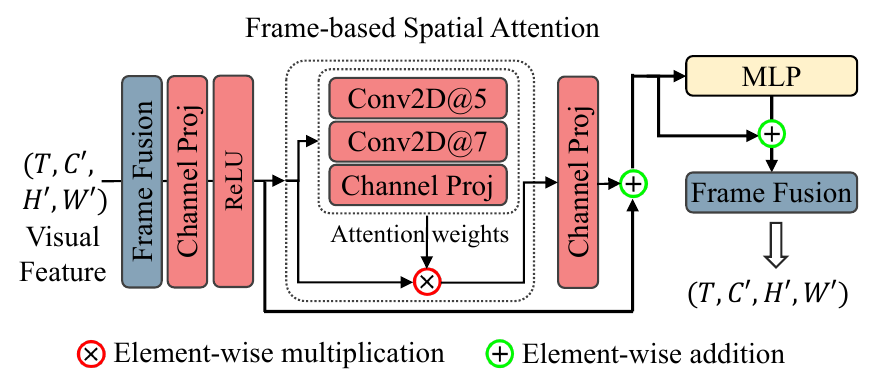}
    \caption{Architecture of the visual preprocessing module.}
    \label{fig:spatial_block}
\end{figure}

\subsection{Multi-modal DeepFake Detection}
\label{Sec.multi-modal_detection}
Recently, many works have utilized the correlations between the audio and visual modalities for video deepfake detection~\cite{mittal2020emotions-EmotionsDontLie,cozzolino_audio-visual_2022-POI,yang_avoid-df_2023_AVoidDF,zhao_self-supervised_2022,chugh2020not-MDS,razaMultimodaltraceDeepfakeDetection2023}. For example, Mittal $et~al.$~\cite{mittal2020emotions-EmotionsDontLie} extracted the emotions cues of the visual and audio modalities, and then utilized the mismatch of their emotions for deepfake detection. 
Zhou $et~al.$~\cite{zhou2021joint-2+1} designed the 2+1-Stream for joint audio-visual detection, which utilizes two streams to detect visual and audio modalities simultaneously and appends a sync stream to exploit the intrinsic synchronization between modalities for the whole video detection. The work~\cite{cheng2022voice_VFD} proposed the VFD that measures the matching degree of face-voice content for deepfake detection. Cai $et~al.$~\cite{cai2022you-LAVDF} proposed a 3D-CNN model for temporal localization of audio and visual manipulations. 

However, existing multi-modal detection methods often learn visual and audio features in isolation, only fusing them during the final classification stage. This strategy may lead to under-utilization of the inherent correlations between audio and visual features. Furthermore, adopting two isolated feature learning sub-models can cause redundant neural layers. As a result, these methods commonly adopt complex network architectures containing numerous parameters. This makes them inefficient and space-consuming and thus unsuitable for practical deployment on resource-limited devices.

\section{Design of SS-AVD}
In this section, we present our SS-AVD in detail. We first describe the overview pipeline of our method, then illustrate the CAVL block, and finally show the multi-modal classification module.

We assume that the shapes of input visual clip $\mathbf{V}$ and audio clip $\mathbf{A}$ are $(T, C, H, W)$ and $(C, L)$, respectively, where $T$ donates the number of video frames, $C$ is the number of channels, $(H, W)$ are the frame height and width, and $L$ indicates the waveform length of audio.

\begin{figure*}[!htbp]
    \centering
    \includegraphics[width=0.99\linewidth]{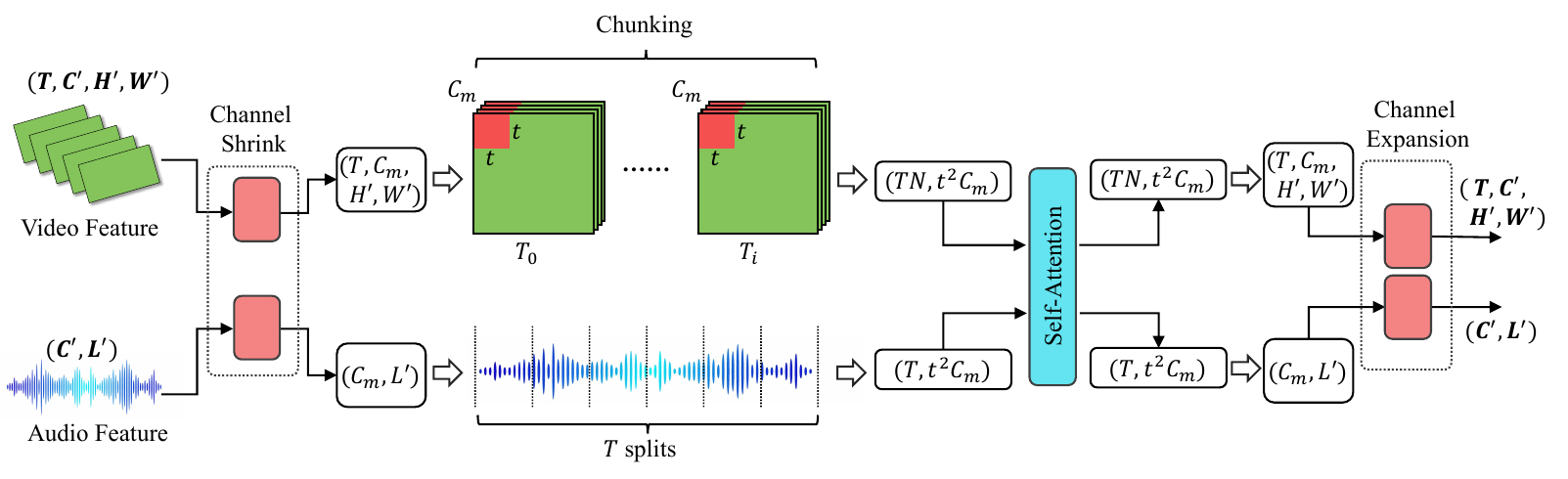}
    \caption{Architecture of the self-attention-based audio-visual module.}
    \label{fig:temporal_block}
\end{figure*}

\subsection{Overview Pipeline}
The overview structure of our SS-AVD is shown in Figure~\ref{fig:architecture}. As can be seen, it is a single-stream network with a classical pyramid structure.
Before fed into each fusion stage, the resolutions of the input audio and visual features are down-sampled, and their channels are increased. In each stage, the input audio and visual features are fed into the stacked CAVL blocks, which perform feature fusion between modalities to learn the audio and visual features collaboratively. Therefore, the visual and audio features are iteratively fused through our single-stream network rather than latent fusion in the final classification. The outputs $\mathbf{F}_{v}$ and $\mathbf{F}_{a}$ of the final fusion stage are then passed a multi-modal classification module for final prediction on the visual, audio, and the whole video.


\subsection{Collaborative Audio-Visual Learning Block}
We propose the CAVL block to learn the visual and audio features collaboratively. Unlike the one-dimensional structure of audio waveform, the visual frames are 2D, and their pixels are spatially correlated. Therefore, the CAVL block first utilizes a VPM to process the visual frames and then employs a SAAVM to fuse visual and audio features. 

\subsubsection{Visual Preprocessing Module} We design the VPM to capture long-distance spatial dependencies for all the visual frames.  The detailed architecture of our VPM is illustrated in Figure~\ref{fig:spatial_block}. Assuming that the input video features is $\mathbf{F}_v'$ with the shape of $(T, C', H', W')$, the VPM first uses a frame fusion layer to fuse the temporal information in the temporal dimension, and then uses frame-based spatial attention to capture the spatial dependencies for visual frames. We implement the frame fusion layer as a single linear layer with weights $\mathbb{R}^{T\times T}$. Given the output $\mathbf{F}_v''$ of the frame fusion, the spatial attention~\cite{guo2022visual-VAN} on each frame is defined as:
\begin{equation}
    \begin{aligned}
        \mathbf{P} &= \operatorname{ReLU} ( \operatorname{Proj}_c(\mathbf{F}_v''),\\
        \mathbf{Q} &= \mathbf{P} \otimes  \operatorname{Proj}_c (\operatorname{DConv}_{7\times7} ( \operatorname{DConv_{5\times5}}(\mathbf{P}))),\\
        Output &=  \mathbf{P} + \operatorname{Proj}_c(\mathbf{Q}),
    \end{aligned}
\end{equation}
where $\operatorname{Proj}_c$ denotes the linear layer on the channel dimension, $\otimes$ means the element-wise multiplication, $\operatorname{DConv}$ represents the depth-wise convolution. Finally, the VPM uses a multilayer perception (MLP) layer to fuse the channel information and a frame fusion layer to further fuse the temporal information. Since spatial attention uses depth-wise convolution layers and shares weights for each frame, our VPM can capture the long-distance dependencies with a small number of parameters.

\subsubsection{Self-Attention Based Audio-Visual Module} We design the SAAVM to learn the spatial-temporal correlation between the visual and audio features. Figure~\ref{fig:temporal_block} shows the architecture of SAAVM, which takes the visual features $\mathbf{F}_v'\in \mathbb{R}^{T\times C' \times H' \times W'}$ and audio features $\mathbf{F}_a'\in \mathbb{R}^{C' \times L'}$ as inputs. 

The SAAVM uses a linear layer to reduce the channel number of the visual and audio features from $C'$ to $C_m$, where $C_m \ll C'$. The reduction of channels can significantly reduce the computational complexity and model parameters in the subsequent attention operations. 
Then, all the $T$ frames of the visual features are chunked with a window of size $t \times t$. By combing all the chunked windows, we can obtain the video tokens $\mathbf{K}_v$ with the shape of $(TN, t^2C_m)$, where $N=\frac{H'W'}{t^2}$. On the other hand, the audio features are split into $T$ splits, and the audio tokens $\mathbf{K}_a \in \mathbb{R}^{T \times t^2C_m}$ are generated by pooling each audio split into a token with the length of $t^2C_m$. 

Next, the SAAVM utilizes a self-attention layer between the visual and audio tokens to capture the spatial-temporal correlations as follows:
\begin{equation}
    \begin{aligned}
        \mathbf{K}_0 &= concat_{d0}(\mathbf{K}_v, \mathbf{K}_a) + PE,\\
        \mathbf{K}_1 &= (softmax(\mathbf{K}_0 \mathbf{K}_0^T/\sqrt{d_k})\mathbf{K}_0)\mathbf{W},\\
        \mathbf{K}_v', \mathbf{K}_a' &= split_{d0}(\mathbf{K}_1),
    \end{aligned}
\end{equation}
where $concat_{d0}$ and $split_{d0}$ indicate the concatenation and split of matrices on the first dimension, respectively, $PE$ denotes the positional embedding~\cite{vaswani2017attention}, $d_k$ indicates the dimension of $\mathbf{K}_1$, and  $\mathbf{W}$ is the weights of a linear layer. The output tokens $(\mathbf{K}_v', \mathbf{K}_a')$ are rearranged and merged into the same shape as the original visual and audio features, respectively. Finally, a linear layer is used to expanse the channel dimension from $C_m$ to $C$. 

\subsection{Multi-Modal Classification Module}
Our SS-AVD first extracts the visual features $\mathbf{F}_{v}\in \mathbb{R}^{T \times C_4 \times H' \times W'}$ and audio features $\mathbf{F}_{a}\in \mathbb{R}^{C_4 \times L'}$ using stacked CAVL blocks with four stages. Then a multi-modal classification module is designed to use these two features to predict the visual label $\hat{y}_{v}$ and audio label $\hat{y}_{a}$, respectively. Besides, we concatenate the visual and audio features to predict the label $\hat{y}_{w}$ for the whole video.
The multi-modal classification module includes a multi-modal style-shuffle augmentation (MMSSA) strategy and a latent-shuffle augmentation (LSA) strategy. The former helps train the detection of the visual and audio modalities, while the latter helps train the detection of the whole video.

\subsubsection{Multi-Modal Style-Shuffle Augmentation} The multimedia content usually has implicit but discriminative styles, such as compression trace, device fingerprint, and background noise. These styles are unrelated to the multimedia content but may be helpful for detection. However, a detection model dependent on these styles will have low evaluation performance when meeting similar content but unseen styles.
Inspired by~\cite{nam2021reducing-domain_gap}, we design the MMSSA strategy to make the visual and audio classifiers focus more on the feature content rather than unrelated styles. Specifically, we regard the mean and variance of the features as the implicit styles~\cite{huang2017arbitrary-style_transfer} and shuffle them between different samples for augmentation. 
The style shuffle of different features is illustrated as follows:
\begin{equation}
        SS(\mathbf{F}^i, \mathbf{F}^j, \omega) =   f(\sigma^i, \sigma^j, \omega) \cdot\left(\frac{\mathbf{F}^i-\mu^i}{\sigma^i}\right)+f(\mu^i, \mu^j, \omega), 
\end{equation}
where $\sigma^*$ and $\mu^*$ denote the mean and variance of the feature $\mathbf{F}^*$, respectively. The $f$ function is used to shuffle styles and is defined as:
\begin{equation}
        f(s^i, s^j, \omega) = \omega \cdot s^i + (1 - \omega) \cdot s^j ,
\end{equation}
where $\omega \in (0,1)$ is to control the shuffle degree. Then, we use two processing heads $H_{v}$ and $H_{a}$ to process the shuffled visual and audio features into vectors, respectively. The $H_{v}$ and $H_{a}$ have similar structures, containing a depth-wise convolutional layer, an adaptive pooling layer, and a linear layer. The prediction process is illustrated as follows:
\begin{equation}
    \begin{aligned}
        \mathbf{Z}_{v}^i, \mathbf{Z}_{a}^i &= H_{v}(SS(\mathbf{F}_{v}^i, \mathbf{F}_{v}^j, \omega)), H_{a}(SS(\mathbf{F}_{a}^i, \mathbf{F}_{a}^j,\omega)), \\
        \hat{\mathbf{y}}_{v}^i, \hat{\mathbf{y}}_{a}^i &= P_v(\mathbf{Z}_{v}^i), P_a({\mathbf{Z}_{a}^i}),
    \end{aligned}
\end{equation}
where $\mathbf{Z}_{v}^i\in \mathbb{R}^{C_4}$ and $\mathbf{Z}_{a}^i\in \mathbb{R}^{C_4}$ are latent features of the visual and audio modalities for the $i$-th sample in the input batch, $P_v$ and $ P_a$ are two linear prediction layers with weights $\mathbb{R}^{C_4\times 2}$,  and $\omega$ is randomly generated in training.

\subsubsection{Latent-Shuffle Augmentation} 
We fuse the latent features of the visual and audio modalities to predict the label of the whole video.
Specifically, we first concatenate the latent features $\mathbf{Z}_{v}^i$ and $ \mathbf{Z}_{a}^i$ on the channel dimension and then use a linear layer for prediction as follows:
\begin{equation}
     \hat{\mathbf{y}}_{w}^i = P_w( concat(\mathbf{Z}_{v}^i, \mathbf{Z}_{a}^i)),
    \label{eq.concat_av}
\end{equation}
where $P_w$ is a linear layer with weights $\mathbb{R}^{2C_4\times 2}$.

    
    

The visual content and audio may be mismatched in the real scenarios due to some permutations, such as environmental noise or recording delay. When a detection model fuses the features from two modalities, the mismatch will cause negative effects on the detection performance. We propose the LSA strategy to reduce the negative effects of this mismatch. Specifically, for each video in the input batch, we randomly combine the visual features and audio features from different samples:
\begin{equation}
    \begin{aligned}
        \tilde{\mathbf{y}}_{w}^i = P_w( concat(\mathbf{Z}_{v}^i, \mathbf{Z}_{a}^j)), \\
    \end{aligned}
\end{equation}
where the ground truth label $\mathbf{y}_w^{i}$ will be changed into zero if  $i\neq j$. During model training, the labels $\tilde{\mathbf{y}}_{w}$ participate in the classification loss to guide the model to enhance the resistance against the mismatches between audio and visual modalities.

\subsection{Loss Function}
The objective loss function of our model consists of classification loss, adversarial loss, and contrast loss.

\subsubsection{Classification Loss} The binary cross-entropy loss (BCE) is commonly used in binary classification tasks to push the predicted probabilities toward the ground truth probabilities:
\begin{equation}
    CE(\hat{\mathbf{y}}, \mathbf{y}) = \frac{1}{N} \sum_{i}^{N}  -(y^{i} \log (\hat{y}^{i})+(1-y^{i}) \log (1-\hat{y}^{i})),
\end{equation}
where $y^{i}$ is the ground truth label, $\hat{y}^{i}$ is the prediction label, and $N$ is the batch size. Since our model can simultaneously predict the labels for the visual, audio, and whole video, the classification loss of our model is defined as:
\begin{equation}
\begin{aligned}
    \mathcal{L}_{cls} = CE(\hat{\mathbf{y}}_v, \mathbf{y}_v) &+ CE(\hat{\mathbf{y}}_a, \mathbf{y}_a) + CE(\hat{\mathbf{y}}_w, \mathbf{y}_w) \\
        & + \beta \cdot CE(\tilde{\mathbf{y}}_w, LSA(\mathbf{y}_w)),\\
\end{aligned}
\end{equation}
where $\beta$ is an adjustment scalar, and $LSA(\mathbf{y}_w)$ indicates changing $\mathbf{y}_w^i$ to zero if the latent features of the visual and audio modalities for the $i$-th sample are shuffled in the LSA strategy.

\subsubsection{Adversarial Loss} Inspired by~\cite{nam2021reducing-domain_gap}, we add the adversarial loss to suppress 
suppress the style-biased representation learning of our network.
Specifically, we append two new processing heads and prediction heads for each modality and obtain two adversarial labels:
\begin{equation}
    \begin{aligned}
        \tilde{\mathbf{y}}_{v}^i &= P_v'(H_{v}'(SS(\mathbf{F}_{v}^j, \mathbf{F}_{v}^i, 0))),\\
        \tilde{\mathbf{y}}_{a}^i &= P_a'(H_{a}'(SS(\mathbf{F}_{a}^j, \mathbf{F}_{a}^i, 0))),
    \end{aligned}
\end{equation}
where $SS(\mathbf{F}_{*}^j, \mathbf{F}_{*}^i, 0)$ means changing the content of $\mathbf{F}_{*}^i$ into $\mathbf{F}_{*}^j$ but maintaining its style. The adversarial loss is calculated as follows:
\begin{equation}
    \mathcal{L}_{adv} = CE(\tilde{\mathbf{y}}_v, \mathbf{y}_{\frac{1}{2}}) + CE(\tilde{\mathbf{y}}_a, \mathbf{y}_{\frac{1}{2}}),
\end{equation}
where $ \mathbf{y}_{\frac{1}{2}}$ indicts a pseudo-label whose elements possess a value of $\frac{1}{2}$ to make equalize the probability of real or fake.

\begin{table*}[!htbp]
      \centering
    \fontsize{8.5pt}{9pt}\selectfont
    \caption{The ACC/AUC scores $(\%)$ of different deepfake detection methods on the evaluation datasets. We report the detection performance on the visual, audio, and whole video. 
Note that the AUC scores for audio manipulation detection are all zeros on DF-TIMIT because it does not contain forged audio, and we report the average performance on the LQ and HQ versions of DF-TIMIT.
    }
    \label{tab.qualitative_res}
  \setlength{\tabcolsep}{1.0mm}{
    
    \begin{tabular}{lrrrrrrrrrrrrrrrrrr}
    \toprule
     \multirow{3}{*}{Method} & \multirow{3}{*}{Param.} & \multicolumn{3}{c}{DF-TIMIT}                   & \multicolumn{3}{c}{FakeAVCeleb}                                               & \multicolumn{3}{c}{DFDC} \\ \cmidrule(lr){3-5}\cmidrule(lr){6-8}\cmidrule(l){9-11}

      & & visual & audio & whole  & visual & audio & whole & visual & audio & whole \\
            \midrule

     2+1 Stream & 64.50M & 94.00/94.00 & \colorblue{100}/0 & \colorred{99.50}/99.50                 & 89.00/95.03& \colorblue{96.44}/98.75 & \colorred{90.00}/95.03                  & 82.92/89.83 & \colorblue{81.62}/84.02 & \colorred{81.74}/89.83                  \\

     Emotions & - & - & - & -\quad/95.60 & - & - & - & - & - & -\quad/84.40\\

     BA-TFD & 5.50M                  & 94.50/94.50 & \colorblue{100}/0 & \colorred{94.50}/94.50                  & 69.22/69.43 & \colorblue{86.66}/83.70 & \colorred{79.77}/78.17                   & 73.33/73.33 & \colorblue{77.22}/53.79 & \colorred{73.25}/73.26                   \\

     VFD & 122.70M  &	- & - & \colorred{99.88}/\quad~~~~- &		- & - & \colorred{81.52}/86.11	 &		- & - & \colorred{80.96}/85.13 \\ 

    MRDF & 20.00M   & - & - & \colorred{95.73}/98.91                  & - & - & \colorred{79.57}/89.35                   &  - & - & 80.27/88.40                   \\
    
     MultiModalTrace & 11.70M & 92.50/92.50 & \colorblue{100}/0 & 92.50/92.50                  & 77.78/78.00 & \colorblue{97.22}/97.88 & \colorred{85.00}/83.72                   & 74.03/80.55 & \colorblue{90.77}/92.87 & \colorred{75.03}/81.56                   \\

                \midrule
     \textbf{SS-AVD} & \textbf{0.48M} & \textbf{97.00}/\textbf{97.00} & \textbf{100}/0 & \textbf{100}~~/\textbf{~~~100}                  & \textbf{98.11}/\textbf{99.37} & \textbf{98.55}/\textbf{99.51} & \textbf{95.11}/\textbf{98.31}                  & \textbf{86.70}/\textbf{93.50} & \textbf{92.33}/\textbf{95.53} & \textbf{86.55}/\textbf{93.61}                  \\

    \bottomrule
    \end{tabular}
    }
\end{table*}

\subsubsection{Contrast Loss} We utilize the contrast loss to maximize the similarity of features with the same labels and minimize the similarity of features with different labels~\cite{chugh2020not-MDS,cai2022you-LAVDF}. Specifically, the contrast loss is defined as follows:
\begin{equation}
    \begin{aligned}
        Contrast(\mathbf{y}, \mathbf{Z})= & \frac{1}{N^2} \sum_i^N\left(\sum_{j: y^i=y^j}^N\left(1-s(\mathbf{Z}^i, \mathbf{Z}^j)\right)\right. \\
        & \left.+\sum_{j: y^i\neq y^j}^N \max \left(s(\mathbf{Z}^i, \mathbf{Z}^j)-\alpha, 0\right)\right),
    \end{aligned}
\end{equation}
where $\mathbf{Z}$ represents the latent features for classification, $s(\mathbf{Z}^i, \mathbf{Z}^j)$ donates the cosine similarity function, 
and $\alpha$ is the margin parameter to control the similarity for label-unmatched sample pairs. The contrast loss of our model is defined as:
\begin{equation}
        \mathcal{L}_{con} = Contrast(\mathbf{y}_v, \mathbf{Z}_{v}) +  Contrast(\mathbf{y}_a, \mathbf{Z}_{a}). 
\end{equation}

The final objective function of our model is weighted from $\mathcal{L}_{cls}$, $\mathcal{L}_{adv}$ and $\mathcal{L}_{contrast}$ as follows:
\begin{equation}
    \mathcal{L} = \gamma_1 \cdot \mathcal{L}_{cls}   + \gamma_2 \cdot \mathcal{L}_{adv}  +  \gamma_3 \cdot \mathcal{L}_{con}, 
\end{equation}
where $\gamma_1$, $\gamma_2$, and $\gamma_3$ are adjusting scalars for each loss.


\section{Experiments}

\subsection{Datasets}
In our experiments, we use three popular deepfake datasets to evaluate our proposed model: 
\begin{itemize}
    \item DF-TIMIT~\cite{korshunov2018deepfakes-DF-TIMIT}: It selected 320 real videos from 32 subjects in the VidTIMIT~\cite{sanderson2002vidtimit} database and used GAN-based approach\footnote{https://github.com/shaoanlu/faceswap-GAN} to produce two versions of video deepfakes. 
    Specifically, it contains 320 low-quality (LQ) deepfakes and 320 high-quality (HQ) deepfakes. Note that the audio track in these video deepfakes was not manipulated. 
    \item DFDC~\cite{dolhansky2020deepfake-DFDC}: It contains more than 100,000 video clips. The videos were produced using several methods, including deepfake generation methods, GAN-based methods, and non-learned methods. Since it only provides video labels, we generate the audio labels by comparing the hash values of audio tracks~\cite{hosler2021deepfakes-dfdc-label}.
    \item FakeAVCeleb~\cite{khalid2021fakeavceleb}: It contains 21544 video clips, in which some deepfake videos have corresponding synthesized lip-synced fake audios. 
\end{itemize}

\textbf{Preprocessing.} We preprocess the dataset to make them suitable for each detection model. For the videos in DFDC and DF-TIMIT, we use a face detection method S3FD~\cite{zhang2017s3fd} to crop and centralize faces. As for videos in FakeAVCeleb, we directly use them since they are already face-centered and cropped~\cite{khalid2021fakeavceleb}. We only use the first three seconds for all videos, where ten frames are extracted even-spaced from the visual clip, and the audio clip is re-sampled with a sample rate of 16 kHz.

\begin{table*}[!t]
    \centering
    
    \caption{The number of deepfakes in different types of used datasets. Considering that the number of Real$_V$-Real$_A$ deepfakes in FakeAVCeleb is unbalanced, we randomly select 2000 real videos from VoxCeleb2~\cite{chung2018voxceleb2} to complement the evaluation set.}
    
    \label{tab.dataset_splits}
    \small
    \begin{tabular}{cccccccccccc}
    \toprule
    \multirow{2}{*}{Type} & \multirow{2}{*}{\tabincell{c}{Visual\\label}} & \multirow{2}{*}{\tabincell{c}{Audio\\label}} &  \multicolumn{2}{c}{FakeAVCeleb} & &\multicolumn{2}{c}{DF-TIMIT} && \multicolumn{2}{c}{DFDC} \\ \cmidrule(lr){4-5} \cmidrule(lr){7-8} \cmidrule(l){10-11}
     & & &  Original & Selected && Original & Selected && Original & Selected \\
    \midrule
    Fake$_V$-Fake$_A$ & 0  & 0 &  10835 & 1500   &    &  0   & 0  &   & 4920  & 4500 \\
    Fake$_V$-Real$_A$ & 0  & 1 &9709  & 1500     &  &  640 & 640   & & 0     & 0\\
    Real$_V$-Fake$_A$ & 1  & 0 &500   & 500      &  &  0   & 0     & & 95072 & 4500\\
    Real$_V$-Real$_A$ & 1  & 1 &500   & 500 + 2000 &  &  320 & 320  & & 19154 & 9000\\
    \midrule
    total           &     &   & 21544  &  4000 + 2000 &  & 960 & 960 && 119146 & 18000 \\
    \bottomrule
    \end{tabular}
\end{table*}

\textbf{Splits.} We randomly select parts of the videos for evaluation. Specifically, we use all videos of DF-TIMIT, randomly select 18000 videos from DFDC, and randomly select 4000 videos from FakeAVCeleb with an extra 2000 real videos from VoxCeleb2~\cite{chung2018voxceleb2} for complementary. Detailed selection is provided in Table\ref{tab.dataset_splits}. As shown in Table.~\ref{tab.dataset_splits}, we divide the deepfakes into four types, Fake$_V$-Fake$_A$, Fake$_V$-Real$_A$, Real$_V$-Fake$_A$, and Real$_V$-Real$_A$, according to the ground truth labels of the visual and audio modalities. We use all the deepfake and real videos in DF-TIMIT and randomly select 18000 videos from DFDC. As for FakeAVCeleb, we randomly select 4000 videos from it and complement it with an extra 2000 real videos from VoxCeleb2~\cite{chung2018voxceleb2}.
We split the training, validation, and test subsets in each evaluation dataset at the rate of 0.75, 0.1, and 0.15, respectively.

\subsection{Experiment Settings}
The shape of the visual frames and audio of the input videos are set as $(3 \times 10 \times 224 \times 224)$ and $(1 \times 48000)$, respectively. The batch size is 32 in training. For the loss function, the margin parameter $\alpha$ is set to 0.4, the $\beta$ is set to 0.5, and the $\gamma_1$, $\gamma_2$, and $\gamma_3$ are set to 1, 0.1, and 1.0, respectively. The $C_m$ in SAAVM is set to 1. The number of fusion stages is set to 4. The numbers $n_i$ of the CAVL blocks in each stage are set to $[2,2,6,2]$, and the channel numbers $C_i$ in these four stages are set to $[8, 16, 32, 64]$.


We train our model for 200 epochs and use data augmentation technologies to enlarge the data diversity, including JPEG compression, flipping, rotating, and Gaussian noise. AdamW optimizer~\cite{loshchilov2017decoupled-AdamW} with a weight decay rate of 0.01 is used to optimize the model parameters. The learning rate is initialized as 0.0005 and then linearly decayed to 0.0001 in the 200 epochs. We implemented all the experiments using the Pytorch framework and on a computer with one RTX4090 GPU device.

\subsection{Comparison Methods}

We compare our method with some state-of-the-art deepfake detection methods. Besides the detection methods 2+1 Stream~\cite{zhou2021joint-2+1}, Emotions~\cite{mittal2020emotions-EmotionsDontLie}, BA-TFD~\cite{cai2022you-LAVDF} and VFD~\cite{cheng2022voice_VFD} introduced in the related work section, we also use the following methods for comparison.
\begin{itemize}
    \item MultiModalTrace~\cite{razaMultimodaltraceDeepfakeDetection2023}: It utilizes two ResNet stems to learn visual and audio features separately and uses an MLP-based fusion module for feature fusion. 
    \item MRDF~\cite{zou2024cross-MRDF}: It employs two feature encoders to extract visual and audio features separably and utilizes a transformer to fuse multi-modal features and make final classification.
 \end{itemize}

Since the methods Emotions and VFD do not provide complete codes, we directly report their test results from their original papers. We implement the 2+1 Stream and MultiModalTrace using the PyTorch framework and use the publicly available codes for the rest of the methods. We train these methods following the settings of their original papers. 

\subsection{Qualitative Results}

To qualitatively analyze our method, we evaluate it on the DF-TIMIT, FakeAVCeleb, and DFDC datasets and compare it with previous methods. Due to the unbalanced distribution of the real and fake samples, we use both the accuracy (ACC) and area under the curve (AUC)~\cite{huang2005using-AUC} to evaluate the performance of all the methods, and Table~\ref{tab.qualitative_res} shows the results. 

\textbf{DF-TIMIT.}  Since the audio tracks in DF-TIMIT are not manipulated, the AUC scores on audio modality are all zeros. For the visual modality and whole video, our SS-AVD can obtain nearly 100$\%$ accuracy and AUC scores on both the LQ and HQ versions. 

\textbf{FakeAVCeleb. } For the audio modality, our SS-AVD can obtain nearly 99$\%$ accuracy and AUC scores, which is superior to those of 2+1 Stream, BA-TFD, and MultiModalTrace methods. For the visual modality, our SS-AVD performs well with accuracy and AUC scores larger than 98$\%$. Considering the detection for the whole video, our SS-AVD outperforms all the other visual-audio joint detection methods. Specifically, our method achieves a 3.28$\%$, 20.14$\%$, 12.2$\%$, 8.96$\%$ and 14.59$\%$ improvement on AUC scores than the 2+1 Stream, BA-TFD, VFD, MRDF and MultiModalTrace, respectively.

\textbf{DFDC.} It can be observed that our SS-AVD outperforms other audio-visual joint detection methods a lot. For example, the AUC score on the audio modality of our SS-AVD is 7.85$\%$ and 0.93$\%$ larger than that of 2+1 Stream and MultiModalTrace, respectively. Compared to all audio-visual joint detection methods, our SS-AVD achieves at least a 1.8$\%$ increase in the AUC scores on detecting the whole video. As for the visual-only detection, our method outperforms all other methods. 


Among all the audio-visual joint detection methods, our SS-AVD can perform the best in nearly all visual, audio, and whole video detection tasks. Moreover, our SS-AVD has only 0.48M parameters, which is much less than other methods. This makes our method highly applicable to resource-limited devices.

\subsection{Cross-method Evaluation}
A robust deepfake detection method should have a high generalization ability on unseen deepfakes. 
We perform the cross-method evaluation on the FakeAVCeleb dataset to evaluate the generalization ability of our method. The FakeAVCeleb dataset is built using two face forgery methods: FaceSwap and Fsgan. We build the training and validation sets using the videos falsified by one method and build the test set using the videos falsified by the other method. Table~\ref{tab.cross_method} shows the test results of cross-method evaluation. Note that Table~\ref{tab.cross_method} does not include the methods Emotions and VFD since there are no corresponding test results in their original papers.
Compared with other visual-audio joint detection methods, our SS-AVD can obtain the best accuracy and AUC scores. This demonstrates that our method has high potential and applicability for real-world deepfake detection.

\begin{table}[!tbp]
      \centering
    \small

    \caption{Cross-method evaluation on FakeAVCeleb, whose deepfake videos are generated by two methods: FaceSwap and Fsgan. The columns ``FaceSwap'' and ``Fsgan'' indicate that the training and validation are run on the deepfake videos generated by them, while the test is performed using the deepfake videos generated by the other method. We report the AUC ($\%$) scores of each method.}

        \label{tab.cross_method}
        
  \setlength{\tabcolsep}{1.25mm}{
    
    \begin{tabular}{lccccccccccc}
    \toprule
     \multirow{4}{*}{Methods} & \multicolumn{6}{c}{FakeAVCeleb}           \\ \cmidrule(l){2-7}
                                                                & \multicolumn{3}{c}{FaceSwap}            & \multicolumn{3}{c}{Fsgan}               \\\cmidrule(lr){2-4}\cmidrule(l){5-7}

     & visual & audio & whole & visual & audio & whole \\ \midrule

 2+1 Stream & 83.87 & \colorblue{99.50} & \colorred{83.61}                   & 81.84 & \colorblue{99.41} & \colorred{84.80} \\ 
                     
    
       BA-TFD & 67.02 & \colorblue{90.83} & \colorred{78.68}                   & 64.05 & \colorblue{88.46} & \colorred{80.56} \\ 

    \tabincell{c}{MRDF} & - & - & \colorred{80.91}                  &  - & - & \colorred{80.00}  \\ 
                     
    \tabincell{c}{MultiModalTrace} & 66.34 & \colorblue{\textbf{99.40}} & \colorred{82.59}                  &  56.04 & \colorblue{99.14} & \colorred{84.62}  \\ 

    \midrule
    \textbf{SS-AVD} & \textbf{85.39} & \colorblue{\textbf{99.84}} & \colorred{\textbf{88.37}}                   & \textbf{88.65} & \colorblue{\textbf{99.91}} & \colorred{\textbf{87.56}} \\ 
    \bottomrule
    \end{tabular}
    }
\end{table}

\section{Ablation Study and Discussion}
We conduct ablation studies on the LSA strategy, the MMSSA strategy, the adversarial loss $\mathcal{L}_{adv}$, and the contrast loss $\mathcal{L}_{con}$ to verify their effectiveness.

\subsection{Ablation Study}

\begin{table}[!ht]
    \centering
    \caption{Ablation studies of the augmentation strategies and loss functions. We report the ACC scores ($\%$) of predicting the whole video on the DFDC test set.}
    \label{tab.ablation_study}
    \small
    \renewcommand{\arraystretch}{0.95}
        \begin{tabular}{cccccccc}
        \toprule
        \multirow{2}{*}{} & \multicolumn{2}{c}{Strategy} &  \multicolumn{2}{c}{Loss}  & \multirow{2}{*}{SS-AVD} \\
        \cmidrule(r){2-3} \cmidrule(r){4-5}
         & LSA & MMSSA & $\mathcal{L}_{adv}$ & $\mathcal{L}_{con}$ &  \\
         \midrule
        (a) & \ding{55} & \ding{55} &  \Checkmark & \Checkmark & 82.96 \\
        (b) & \ding{55} & \Checkmark &  \Checkmark & \Checkmark &85.25 \\
        (c) & \Checkmark & \ding{55} &  \Checkmark & \Checkmark & 83.33 \\
        (d) & \Checkmark &  \Checkmark & \ding{55} & \Checkmark & 85.14 \\
        (e) & \Checkmark & \Checkmark &  \Checkmark & \ding{55} & 86.03 \\
        (f) & \Checkmark & \Checkmark &  \Checkmark & \Checkmark & 86.48  \\
        \bottomrule
        \end{tabular}
\end{table}

\textbf{Augmentation strategies.} The LSA strategy reduces the negative impact of possible mismatches between the visual and audio modalities, while the MMSSA strategy makes the modality classifiers focus on the feature content rather than unrelated styles. To evaluate the effectiveness of these two strategies, we conduct ablation studies on them, and Table~\ref{tab.ablation_study} shows the results on the test set of DFDC. As can be seen from the results of the settings (a), (b), (c), and (f) in Table~\ref{tab.ablation_study}, both the LSA and MMSSA strategies can improve the model performance.

\textbf{Loss functions.} We add the adversarial loss  $\mathcal{L}_{adv}$ to suppress the detection accuracy on the feature styles and the contrast loss $\mathcal{L}_{con}$  to increase the feature discriminability. As shown from the results of the setting (d), (e) and (f) in Table~\ref{tab.ablation_study}, the ablation studies on them indicate that both the adversarial and contrast losses benefit the model performance. Specifically, the $\mathcal{L}_{adv}$ and $\mathcal{L}_{con}$ losses improve the ACC scores by 1.34$\%$ and 0.45$\%$ for our SS-AVD, respectively.

\textbf{Hyperparameters}
We fine-tune the hyperparameters $\{\gamma_1, \gamma_2, \gamma_3\}$ to assess the sensitivity of the model's performance to them. Since our method entails a classification model, we maintain the weight $\gamma_1$ of the classification loss fixed at 1.0 and vary only the other two weights $\{\gamma_2, \gamma_3\}$. 
\begin{table}[!h]
    \centering
    \small
    \caption{ACC ($\%$) scores of entire video detection on DFDC.}
    \begin{tabular}{cccccc}
    \toprule
        $(\gamma_2$,$\gamma_3$) & (0.2, 1.0) & (0.5, 1.0) & (1.0,1.0) \\ 
        ACC &  84.35 & 86.01 & 85.35 \\ \hline
        $(\gamma_2$,$\gamma_3$) & (0.1, 0.5) & (0.1, 0.2) & \textbf{(0.1,1.0)}\\ 
        ACC & 82.79 & 84.70 & \textbf{86.48}\\
    \bottomrule
    \end{tabular}
    \label{tab:para}
\end{table}
As illustrated in Table~\ref{tab:para}, our method exhibits sensitivity to these hyperparameters. Due to constraints in computing resources, we tested only a limited set of combinations. While our default settings ($\{0.1,1.0\}$) may not be optimal, they indicate that the performance of our method can be enhanced further through improved hyperparameter combinations.

\subsection{Model Discussion}

\begin{figure*}[t]
  \centering
    \begin{minipage}[b]{0.8\linewidth}
        \centering
        \begin{minipage}[b]{0.12\linewidth}
            \vspace{1cm}
            \leftline{\text{(a) Frame 1}}
            \vspace{1cm}
        \end{minipage}
        \begin{minipage}[b]{0.17\linewidth}
            \centerline{\includegraphics[width=1\linewidth]{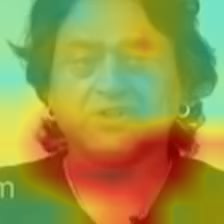}}
        \end{minipage}
        \begin{minipage}[b]{0.17\linewidth}
            \centerline{\includegraphics[width=1\linewidth]{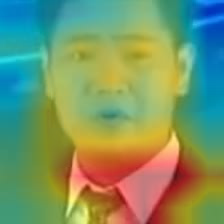}}
        \end{minipage}
        \begin{minipage}[b]{0.17\linewidth}
            \centerline{\includegraphics[width=1\linewidth]{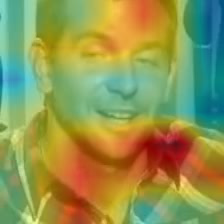}}
        \end{minipage}
        \begin{minipage}[b]{0.17\linewidth}
            \centerline{\includegraphics[width=1\linewidth]{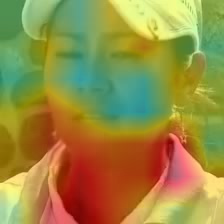}}
        \end{minipage}
        \begin{minipage}[b]{0.17\linewidth}
            \centerline{\includegraphics[width=1\linewidth]{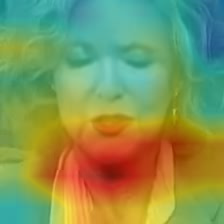}}
        \end{minipage}
    \end{minipage}

    \begin{minipage}[b]{0.8\linewidth}
        \centering
        \begin{minipage}[b]{0.12\linewidth}
            \vspace{1cm}
            \leftline{\text{(b) Frame 6}}
            \vspace{1cm}
        \end{minipage}
        \begin{minipage}[b]{0.17\linewidth}
            \centerline{\includegraphics[width=1\linewidth]{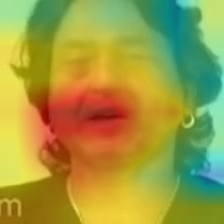}}
        \end{minipage}
        \begin{minipage}[b]{0.17\linewidth}
            \centerline{\includegraphics[width=1\linewidth]{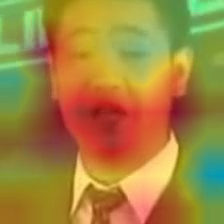}}
        \end{minipage}
        \begin{minipage}[b]{0.17\linewidth}
            \centerline{\includegraphics[width=1\linewidth]{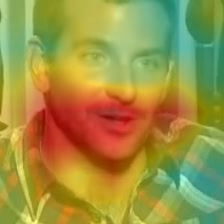}}
        \end{minipage}
        \begin{minipage}[b]{0.17\linewidth}
            \centerline{\includegraphics[width=1\linewidth]{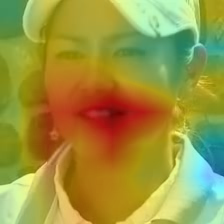}}
        \end{minipage}
        \begin{minipage}[b]{0.17\linewidth}
            \centerline{\includegraphics[width=1\linewidth]{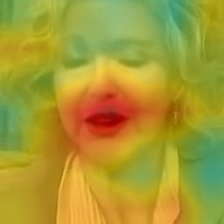}}
        \end{minipage}
    \end{minipage}

    \begin{minipage}[b]{0.8\linewidth}
        \centering
        \begin{minipage}[b]{0.12\linewidth}
            \vspace{1cm}
            \leftline{\text{(c) Frame 10}}
            \vspace{1cm}
        \end{minipage}
        \begin{minipage}[b]{0.17\linewidth}
            \centerline{\includegraphics[width=1\linewidth]{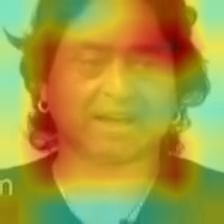}}
        \end{minipage}
        \begin{minipage}[b]{0.17\linewidth}
            \centerline{\includegraphics[width=1\linewidth]{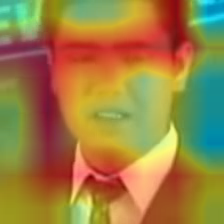}}
        \end{minipage}
        \begin{minipage}[b]{0.17\linewidth}
            \centerline{\includegraphics[width=1\linewidth]{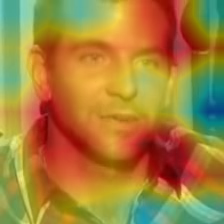}}
        \end{minipage}
        \begin{minipage}[b]{0.17\linewidth}
            \centerline{\includegraphics[width=1\linewidth]{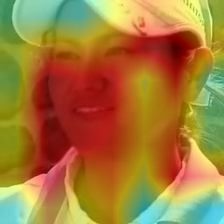}}
        \end{minipage}
        \begin{minipage}[b]{0.17\linewidth}
            \centerline{\includegraphics[width=1\linewidth]{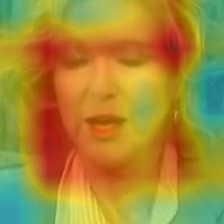}}
        \end{minipage}
    \end{minipage}

    \caption{Visualization of the regions focused on by our SS-VAD for prediction. The figures from the top to bottom rows are the first, sixth, and tenth frames of the input videos, respectively.}
    \label{Fig.grad_cam}
\end{figure*}

\subsubsection{Visualization.} 
We visually investigate the regions focused on by our SS-VAD to detect the whole video. Specifically, we employ GradCAM~\cite{selvaraju2017gradcam} on the video features $\mathbf{F}_v$ to highlight the most relevant regions of each visual frame for the classification decision. We utilize the test set of FakeAVCeleb for presentation. Figure~\ref{Fig.grad_cam} plots the heatmaps where a region with a warmer color is more important to the prediction. As seen from Figure~\ref{Fig.grad_cam}, our SS-VAD focuses on different areas of the frame, such as the neck, mouth, and face area. These areas commonly exist visual artifacts for detecting visual frames~\cite{huang2021deepfake-deepfakeMNIST}. Besides, since lip-syncing is used to synchronize the visual frames and the audio deepfake, these areas also correlate to the synchronization between the visual and audio modalities. Therefore, our SS-VAD can effectively capture the correlation between different modalities and find the visual artifacts to make decisions jointly.

\subsubsection{Feature representations.} 

\begin{figure*}[!ht]
  \centering
    \begin{minipage}[b]{0.99\linewidth}
        \centering
        \begin{minipage}[b]{0.19\linewidth}
            \centerline{\includegraphics[width=1\linewidth]{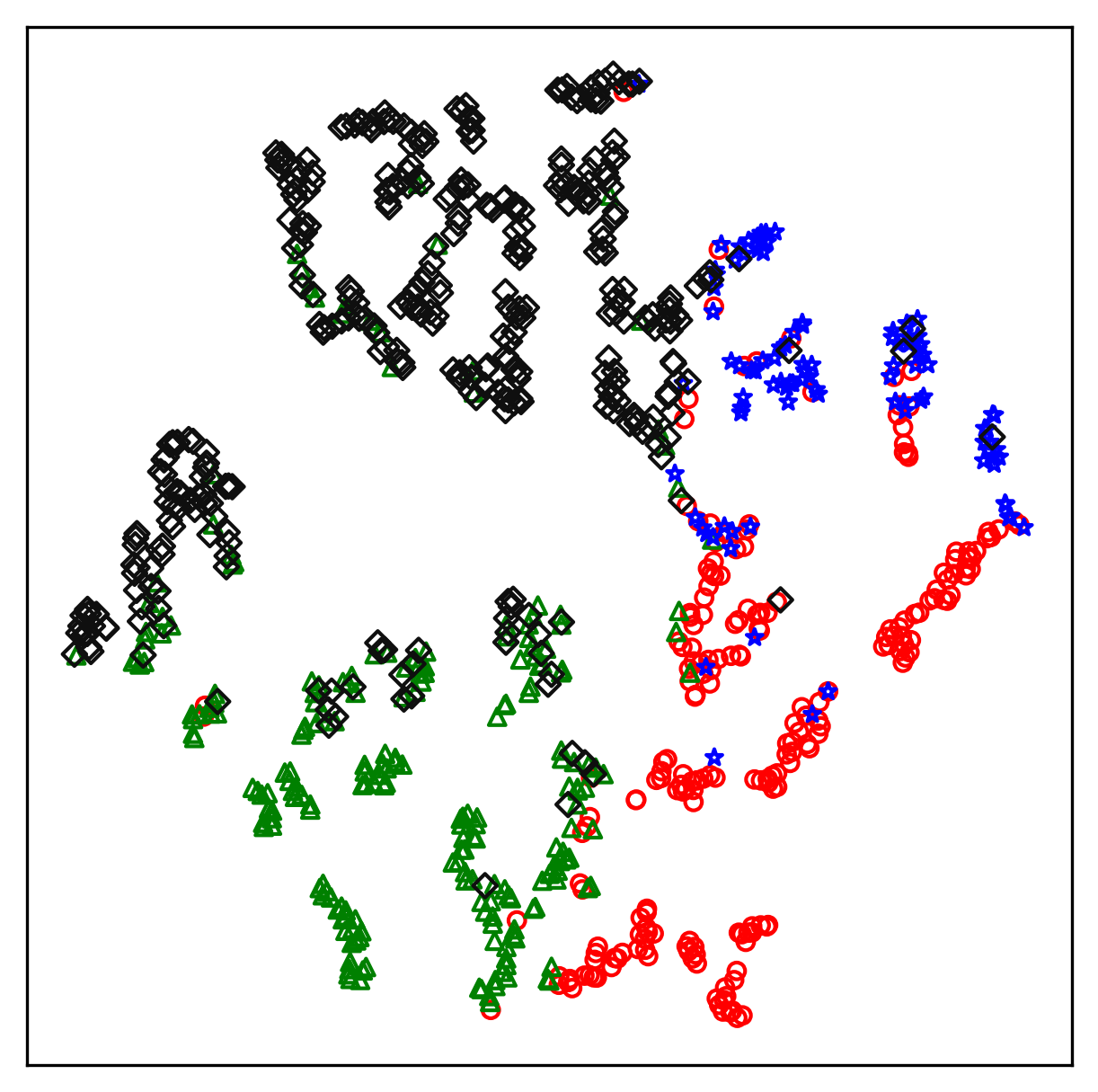}}
            \centerline{2+1}
        \end{minipage}
        \centering
        \begin{minipage}[b]{0.19\linewidth}
            \centerline{\includegraphics[width=1\linewidth]{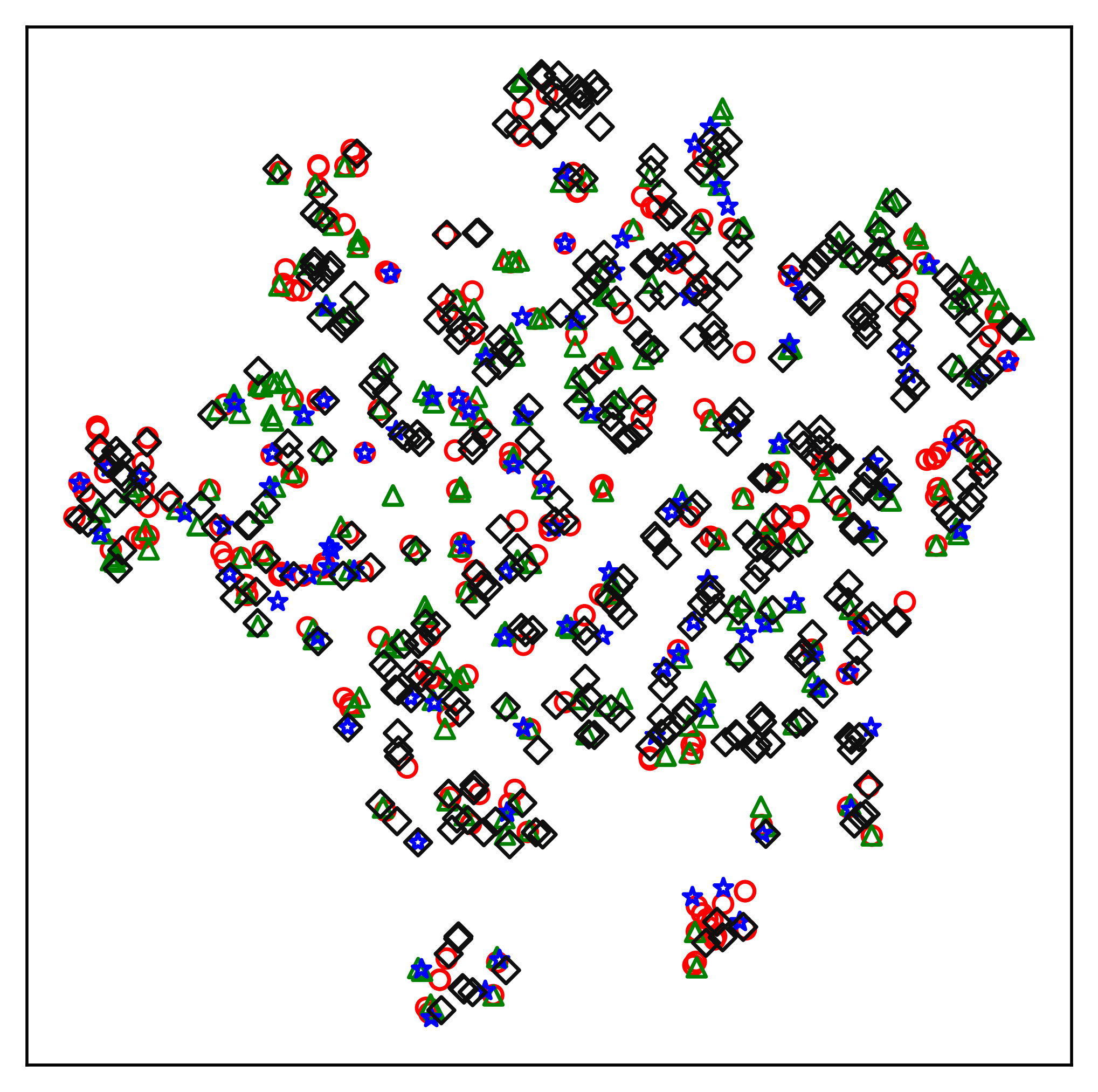}}
            \centerline{BA-TFD}
        \end{minipage}
        \begin{minipage}[b]{0.19\linewidth}
            \centerline{\includegraphics[width=1\linewidth]{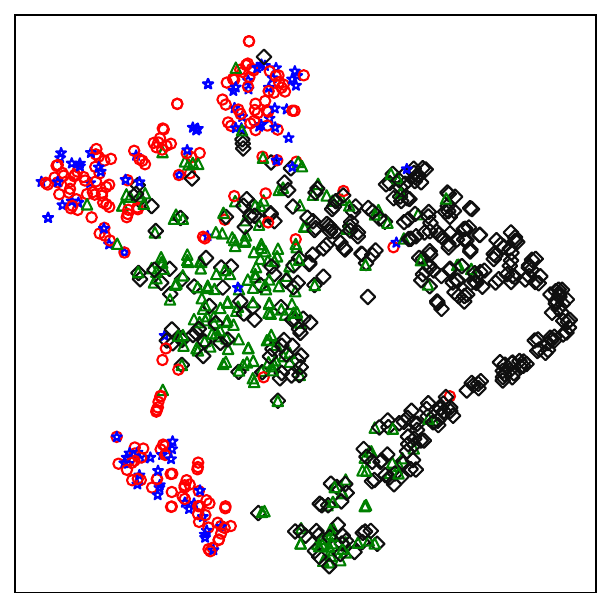}}
            \centerline{MRDF}
        \end{minipage}
        \begin{minipage}[b]{0.19\linewidth}
            \centerline{\includegraphics[width=1\linewidth]{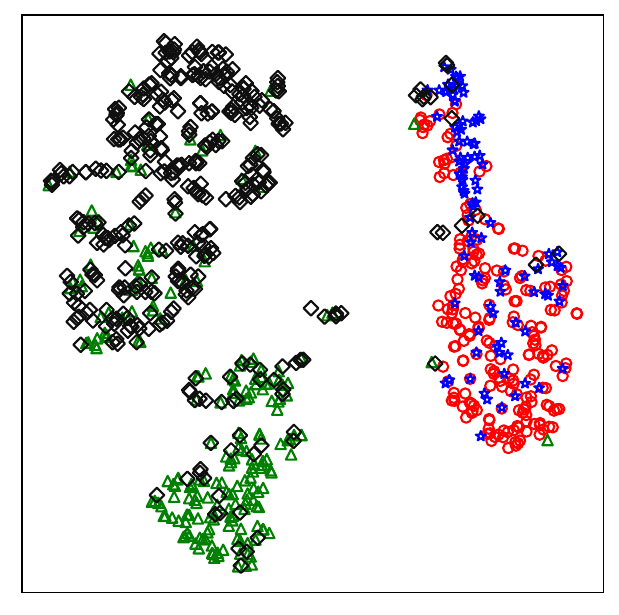}}
            \centerline{MultiModalTrace}
        \end{minipage}
        \begin{minipage}[b]{0.19\linewidth}
            \centerline{\includegraphics[width=1\linewidth]{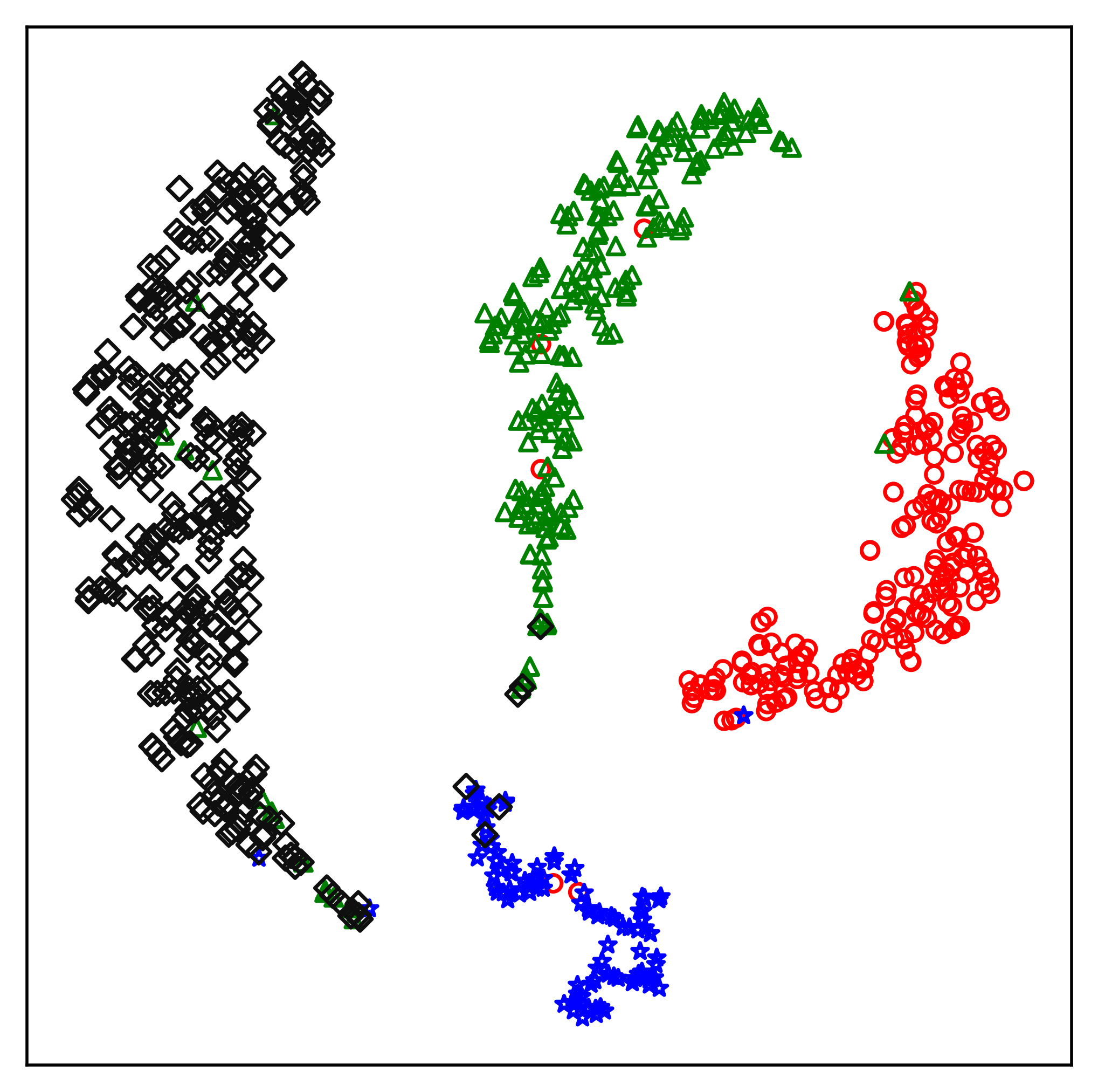}}
            \centerline{\textbf{SS-AVD}}
        \end{minipage}
    \end{minipage}

    \vspace{5pt}
    \begin{minipage}[b]{0.49\linewidth}
        \centering
        \begin{minipage}[b]{0.9\linewidth}
            \centerline{
              \begin{tcolorbox}[colback=white,colframe=black!75!white,arc=1mm, boxsep=-1mm, boxrule=0.04mm]
                \begin{center}
                \includegraphics[width=0.95\linewidth]{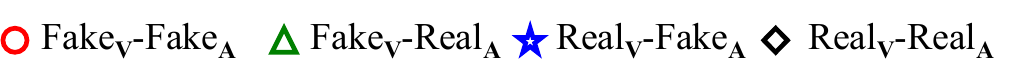}
                \end{center}
              \end{tcolorbox}
            }
        \end{minipage}
    \end{minipage}
    \caption{t-SNE results of the latent features on the test set of FakeAVCeleb.}
    \label{Fig.t_SNE}
\end{figure*}

To intuitively show the effectiveness of the feature learning, we utilize t-SNE~\cite{arora2018analysis-t-sne} to visualize the feature clustering for four types of deepfakes: Fake$_V$-Fake$_A$, Fake$_V$-Real$_A$, Real$_V$-Fake$_A$, and Real$_V$-Real$_A$. The test set of FakeAVCeleb is also used for the presentation. Specifically, the latent features of the whole video is used for clustering for our SS-AVD, the 2+1~\cite{zhou2021joint-2+1}, MRDF~\cite{zou2024cross-MRDF} and MultiModalTrace~\cite{razaMultimodaltraceDeepfakeDetection2023}. For the BA-TFD~\cite{cai2022you-LAVDF}, we add the latent features of visual and audio modalities for clustering since it does not have the latent features of the whole video. Figure~\ref{Fig.t_SNE} shows the clustering results. One can see that our SS-AVD has the best discriminating features. This indicates that our method has better feature learning ability.

\section{Conclusion}
This paper presents SS-AVD, a lightweight single-stream network for joint audio-visual deepfake detection. We designed a CAVL block to learn the visual and audio features collaboratively. By iteratively employing this block, our network continuously fuses multi-modal features across its layers.
Besides, we propose a multi-modal classification module to predict the visual, audio, and whole video labels. The multi-modal classification module employs MMSSA and LSA strategies to assist training. We evaluate our method on three audio-visual benchmark datasets, DF-TIMIT, FakeAVCeleb, and DFDC. The experimental results demonstrate that our method outperforms the state-of-the-art audio-visual joint detection methods while maintaining the minimal number of parameters.

\bibliography{aaai25}

\end{document}